\documentclass[pdflatex,sn-mathphys-num]{sn-jnl}

\usepackage{graphicx}%
\usepackage{multirow}%
\usepackage{amsmath,amssymb,amsfonts}%
\usepackage{amsthm}%
\usepackage{mathrsfs}%
\usepackage[title]{appendix}%
\usepackage{xcolor}%
\usepackage{textcomp}%
\usepackage{manyfoot}%
\usepackage{booktabs}%
\usepackage{algorithm}%
\usepackage{algorithmicx}%
\usepackage{algpseudocode}%
\usepackage{listings}%
\usepackage{bm}

\theoremstyle{thmstyletwo}%

\theoremstyle{thmstylethree}%

\begin{document}

\title[Article Title]{Learning Non-Local Molecular Interactions via Equivariant Local Representations and Charge Equilibration}


\author[1]{\fnm{Paul} \sur{Fuchs}}

\author[1]{\fnm{Micha\l} \sur{Sanocki}}

\author*[1,2]{\fnm{Julija} \sur{Zavadlav}}\email{julija.zavadlav@tum.de}

\affil[1]{\orgdiv{Multiscale Modeling of Fluid Materials, Department of Engineering Physics and
Computation}, \orgname{TUM School of Engineering and Design, Technical University of Munich}, \country{Germany}}

\affil[2]{\orgdiv{Atomistic Modeling Center, Munich Data Science Institute}, \orgname{Technical University of
Munich}, \country{Germany}}


\abstract{Graph Neural Network (GNN) potentials relying on chemical locality offer near-quantum mechanical accuracy at significantly reduced computational costs.
Message-passing GNNs model interactions beyond their immediate neighborhood by propagating local information between neighboring particles while remaining effectively local.
However, locality precludes modeling long-range effects critical to many real-world systems, such as charge transfer, electrostatic interactions, and dispersion effects.
In this work, we propose the Charge Equilibration Layer for Long-range Interactions (CELLI) to address the challenge of efficiently modeling non-local interactions. This novel architecture generalizes the classical charge equilibration (Qeq) method to a model-agnostic building block for modern equivariant GNN potentials.
Therefore, CELLI extends the capability of GNNs to model long-range interactions while providing high interpretability through explicitly modeled charges.
On benchmark systems, CELLI achieves state-of-the-art results for strictly local models.
CELLI generalizes to diverse datasets and large structures while providing high computational efficiency and robust predictions.}

\maketitle

\section{Introduction}

\label{Introduction}

Machine learning potentials (MLPs) are powerful tools for modeling interatomic interactions.
They can achieve near-quantum mechanical accuracy at a fraction of the computational cost~\cite{Anstine2023} and linear scaling with the number of particles.
Thus, highly scalable and accurate MLPs can enable precise simulations of larger systems and allow computational studies of complex phenomena that would otherwise be computationally inaccessible~\cite{frank2024euclideanfastattentionmachine,  Ko2021,Rocken2024, Rocken2024_2}.
Particularly, equivariant Graph Neural Network (GNN) MLPs such as Allegro~\cite{Musaelian2023} and MACE~\cite{NEURIPS2022MACE} are highly expressive models and can learn potential energy surfaces end-to-end from data~\cite{Batzner2022}.
Therefore, these models generalize well even for chemically highly diverse datasets~\cite{kovacsMACEOFFShortRangeTransferable2025,Yang2025-ct}.
However, strictly local MLPs, which assume that atomic interactions are dominated by their immediate environment~\cite{Anstine2023}, and message-passing MLPs, which propagate information beyond the immediate environment~\cite{gilmerMPGNN}, cannot model interactions beyond a strict effective cutoff radius~\cite{kosmala_ewaldbased_2023}.
Thus, these effectively local MLPs can accurately capture short-range interactions~\cite{Musaelian2023}, but cannot capture long-range electrostatic interactions, charge transfer, and dispersion effects~\cite{Ko2021, Shaidu2024}.

Without additional mechanisms to address long-range interactions,  the effective locality of most MLPs greatly limits their application to many real-world scenarios~\cite{Anstine2023,unkeMachineLearningForce2021a}. Long-range interactions are crucial in several key physical phenomena, including molecular aggregation, protein folding, or the behavior of ionic liquids~\cite{protein2, BASKIN2022118616}.
For instance, in proteins, long-range interactions have been shown to play a significant role in their structure and function, with most residues participating in such interactions~\cite{protein, protein2}, some of which can theoretically span up to 15 \AA\;\cite{PhysRevResearch.5.L012028}. As a result, even MLPs with near quantum-level accuracy for short-range interactions would be unable to fully capture the behavior of such proteins without additional schemes considering these long-range effects.

The challenge of modeling long-range effects is a recognized obstacle in developing MLPs~\cite{Anstine2023,frank2023so3kratesequivariantattentioninteractions}.
Thus, several approaches to incorporate long-range effects into GNN MLPs have been proposed.
Reciprocal space methods model long-range interactions by processing structural~\cite{yu2022capturinglongrangeinteractionreciprocal} or learned features~\cite{kosmala_ewaldbased_2023,wang2024neuralp3mlongrangeinteracti} in Fourier space.
However, these methods based on lattice vectors have limited generalizability, as they struggle with differently oriented structures or other supercells and cannot easily be applied to simulations in realistic conditions \cite{frank2024euclideanfastattentionmachine}.
The Euclidean Fast Attention (EFA) scheme~\cite{frank2024euclideanfastattentionmachine} overcomes this limitation and respects relevant physical symmetries but requires integrating possible lattice orientations over the unit sphere, increasing computational costs.
Methods such as Long-Short-Range Message-Passing, RANGE, and Erwin aim to model long-range interactions by improving the efficiency of message passing by utilizing coarse-grained or hierarchical representations \cite{li2024longshortrangemessagepassingphysicsinformedframework, caruso2025extendingrangegraphneural, zhdanov2025erwintreebasedhierarchicaltransformer}.
However, these methods are not directly generalizable to bulk systems.

On the other hand, physics-driven approaches have been proposed.
The simplest approaches treat short-range interactions with MLPs separately from long-range interactions.
Therefore, long-range contributions, such as van der Waals~\cite{10.1063/5.0005084} or electrostatic interactions~\cite{PhysRevB.83.153101, Coste2023}, are subtracted from MLP training data and added to MLP predictions.
However, long-range interactions often correlate with the immediate environments of atoms.
For example, capturing long-range interactions through electrostatic effects requires atomic charges that depend on the dynamic chemical environment.
Therefore, methods have since emerged that predict charges~\cite{Song2024, D3SC02581K} or directly model long-range interactions~\cite{Kim2024LearningCA}, using features of the local environment.
Still, these point charge-based methods generally assume chemical locality, which becomes problematic in systems where non-local effects dominate~\cite{Unke2021}.
Moreover, direct charge prediction often requires additional correction schemes to ensure charge conservation and prevent unphysical behaviour~\cite{Song2024, D3SC02581K}.
Thus, traditional methods fail to account for, e.g., local bonding environments and the global electrostatic landscape~\cite{CENT}.
The Charge Equilibration Neural Network (CENT)~\cite{CENT} was introduced by \citeauthor{CENT} and later adapted by \citeauthor{Ko2021} and \citeauthor{Shaidu2024} to address these challenges in coupling long-range and short-range effects.
The CENT method globally distributes charges via the Charge Equilibration method (Qeq) based on electrostatic features of the local environment extracted via a Behler-type neural network~\cite{PhysRevB.83.153101}.
Therefore, the CENT method correlates short- and long-range effects.
Moreover, explicitly predicting charge distributions can be advantageous, as charge transfer can be observed, and systems in external electric fields can be simulated using predicted charges.
Still, the CENT method does not explicitly account for short-range non-electrostatic interactions.
To overcome this issue, the fourth-generation high-dimensional neural network potentials (4GHDNNPs) comprise a second neural network, modeling short-ranged interactions dependent on the charge state~\cite{Ko2021}.
This method accurately captures global charge distributions in simple systems with non-local effects but requires training an additional neural network on ambiguously defined reference point charges~\cite{Ko2021,Shaidu2024}.
Thus, alternative methods predict electrostatic features alongside the short-range corrections using a single Behler-type neural network and a more fidelity Qeq scheme without reference charges~\cite{Shaidu2024} or replace the Qeq method with a self-consistent method to represent electrostatic interactions using well-defined Maximally Localized Wannier Function Centers~\cite{SCFNN}.
Still, these methods require multi-step training procedures and employ Behler-type neural networks relying on hand-crafted descriptors.
Therefore, they are not simple to generalize to chemically diverse datasets~\cite{Batzner2022}.

Previous machine-learning approaches are often costly, hard to scale, or not simple to generalize to chemically diverse systems.
Thus, this work introduces the Charge Equilibration Layer for Long-range Interactions (CELLI), a novel architectural building block for equivariant GNN MLPs.
By generalizing the Qeq method to chemically diverse systems, CELLI enables MLPs to model long-range interactions and condition short-range interactions on the local charge environment via learned representations.
In a series of experiments with crucial charge-transfer and charge-state dependence~\cite{Ko2021}, we show that integrating CELLI with Allegro~\cite{Musaelian2023} and MACE~\cite{NEURIPS2022MACE} can overcome the inherent locality of state-of-the-art MLPs.
Moreover, on the OE62 dataset~\cite{OE62}, we demonstrate that CELLI can generalize across a more diverse chemical space while only marginally increasing computational costs.
In addition, we employ CELLI on subsets of the SPICE dataset~\cite{eastmanSPICEDatasetDruglike2023} to prove that it can produce stable molecular dynamics simulations.

\section{Results}

\paragraph{Charge Equilibration Layer for Long-range Interactions (CELLI)}

With the Charge Equilibration Layer for Long-range Interactions (CELLI), we introduce the classical non-local Qeq method to recent expressive equivariant GNN architectures.
Similar to \citeauthor{Ko2021}, we split the total potential energy
$U = U_\text{Coul} + \Delta U$
into an electrostatic component $U_\text{Coul}$ and a correction $\Delta U$.
Following the CENT approach~\cite{CENT}, CELLI leverages the Qeq method, accounting for non-local charge transfer, to compute partial charges $\bm Q$ and electrostatic energy $U_\text{Coul}$ (subsection~\ref{subsec:qeq_method}) using features of the GNN.
Subsequently, the GNN learns the correction $\Delta U$ dependent on non-local features provided through the equilibrated partial charges embedded by CELLI.
Thus, instead of learning charges and the potential energy through separate NNs~\cite{Ko2021}, CELLI enables flexible integration of Qeq-based charge prediction and non-local interaction modeling within a single model.

\begin{figure}
\begin{minipage}{.67\linewidth}
    \includegraphics[width=\linewidth]{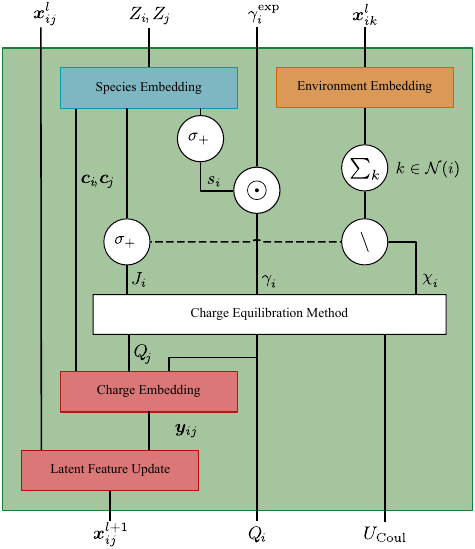}
\end{minipage}\hfill
\begin{minipage}{.285\linewidth}
    \vskip .5em
    \includegraphics[width=\linewidth]{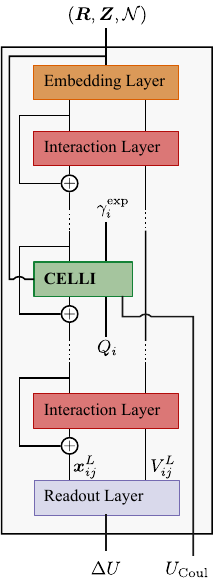}  
\end{minipage}
\caption{\textbf{Charge Equilibration Layer for Long-range Interactions (CELLI).}
\textbf{Left:}
CELLI updates scalar latent features $\bm x^l_{ij}$ from the previous layer $l$ using node species $Z_i$ and covalent radii $\gamma_i$. $\sigma_+$ denotes the generalized softplus function, $\odot$ an elementwise multiplcation, $\Sigma_k$ the sum over all incident edges, and $\backslash$ a split in the feature dimension. The dashed line denotes an optional connection.
\textbf{Right:} CELLI included in the strictly local Allegro architecture. Allegro predicts the total energy $U = U_\text{Coul} + \Delta U$ and partial charges $Q$ using scalar $\bm x_{ij}^l$ and tensorial $V_{ij}^l$ edge features of an input graph with node positions $\bm R$, node species $\bm Z$, and connectivity $\mathcal N$. $\oplus$ denotes a weighted residual update. Radii $\gamma_i^\text{exp}$ are only used by CELLI.}  \label{fig:AllegroQeq}    \label{fig:CELLI}
\end{figure}

We describe CELLI using the example of the Allegro architecture~\cite{Musaelian2023}, visualized in Figure~\ref{fig:AllegroQeq}.
Allegro is a strictly local equivariant GNN that learns scalar features $\bm x_{ij}^l$ and tensorial features $V_{ij}$ for the directed edges $ij$ of a graph through a sequence of $L$ tensor-product layers (outlined in subsection~\ref{subsec:gnns}), which we call \emph{Interaction Layers} in the following.

To extend Allegro, we insert one instance of CELLI at a location between the Interaction Layers.
CELLI embeds and applies the Qeq method to learn long-range electrostatic interactions and partial charges using the latent scalar features.
As global charge transfer might affect the local electronic structure and thus the local many-body interactions~\cite{Ko2021}, CELLI embeds the charge environment into the latent features $\bm x_{ij}^{l+1}$ passed to the following tensor product layers. Finally, the readout layer uses the latent features to predict per-edge energies, summing up to the correction potential $\Delta U$.

\begin{description}
\item[Environment Embedding]~\\
The latent features from the previous $l$ Interaction Layers encode scalar descriptions of the local edge environments.
We use a multi-layer perceptron $\operatorname{MLP}_{\mathcal R}$ to predict a per-edge contribution to the electronegativity, and optionally the hardness $\left(\tilde \chi_{ij}, \tilde J_{ij}^\mathcal R\right) = \operatorname{MLP}_{\mathcal R}\left(\bm x_{ik}^{l}\right)$.
Summing up the contributions of all directed edges from $i$ to $j$, multiplied by a species-invariant learnable factor $f$, yields the electronegativity $\chi_i = f\sum_{k\in \mathcal N(i)} \tilde \chi_{ij}$ of particle $i$.\\

\item[Species Embedding]~\\
We encode the particle species to an environmentally independent contribution to the hardnesses $\tilde J_i^Z$.
To ensure positive hardnesses, we use the generalized soft plus activation function denoted as $\sigma_+(x_1, \ldots, x_k) = \log(1 + \exp(x_1) + \ldots + \exp(x_k))$ to combine the species-dependent with the environment-dependent hardness contributions.
Therefore, we obtain the particle hardness $J_i = \sigma_+\left(\tilde J_i^Z, \sum_{j \in \mathcal N(i)} \tilde J_{ij}^\mathcal R\right)$.
Appropriate radii can crucially determine whether the optimization converges.
Therefore, we base the charge radii on single-bond covalent radii $\gamma_i$~\cite{pyykkoMolecularSingleBondCovalent2009} and learn a positive species-dependent scaling factor $\tilde s_i$ to obtain $\gamma_i = \frac{\sigma_+(\tilde s_i)}{\log(2)}\gamma_i^\text{exp}$.
Additionally, we embed species as features $\bm c_i(Z_i)$ for later use in the charge~embedding.\\

\item[Charge Equilibration Method (Qeq)]~\\
The charge equilibration method takes the environment-dependent electronegativities, species or environment-dependent hardnesses, and species-dependent charge radii to predict partial charges and the coulombic potential (subsection~\ref{subsec:qeq_method}).
As different systems require different treatments of the long-range electrostatic interactions, the method optionally employs, e.g., the Smooth Particle Mesh method~\cite{essmannSmoothParticleMesh1995} for periodic systems.\\

\item[Charges Embedding and Latent Feature Update]~\\
We embed the equilibrated charge environment into the scalar features to provide non-local information to the network.
Therefore, a first multi-layer perceptron $\operatorname{MLP}_Q$ generates charge-dependent features $\bm y_{ij} = \operatorname{MLP}_Q(Q_i, Q_j, \bm c_i, \bm c_j)$ using the equilibrated charges and species embedding $\bm c_i$ of the central and neighbor atoms. These charge dependent features $\bm y_{ij}$ are used by a second multi-layer-perceptron $\operatorname{MLP}_{\bm x}$ to update the scalar features from the previous layer $\bm x_{ij}^l$ to $\bm x_{ij}^{l+1} = \operatorname{MLP}_{\bm x}\left(\bm y_{ij}, \bm x_{ij}^{l}\right)p_\text{env}\left(\lVert \bm R_i - \bm R_j \rVert\right)$, where $p_\text{env}$ is a polynomial envelope function.

\end{description}

\paragraph{Benchmark systems with strictly local models}

First, we test our approach on four benchmark systems introduced by \citeauthor{Ko2021}.
These systems were constructed to be unsolvable by strictly local and charge-independent methods, and allow for visual inspection to verify that the model does not exhibit unphysical behavior, and include: Carbon Chains, Silver Clusters , Sodium Chloride Clusters and Gold Dimers on MgO(001) surface (Methods, Section~\ref{subsec:systems}). They have also been used in related works, enabling a direct comparison with other approaches based on strictly local models~\cite{Ko2021, Shaidu2024, Kim2024LearningCA}.

\begin{table*}[t]
    \centering
    \vskip -0.15in
    \scriptsize
    \caption{Root mean square errors (RMSE) in units of meV/atom, meV/\AA, and me, for CELLI applied to strictly local Allegro model in comparison to the baseline Allegro model and the previous local descriptor methods 4G~\cite{Ko2021}, LRSR~\cite{Shaidu2024}, and CACE-LR~\cite{Kim2024LearningCA} modeling long-range interactions. $J^\mathcal{R}$ denotes CELLI with environment-dependent hardness instead of purely species-dependent hardness.
Errors for models with a larger cutoff than in the original reference~\cite{Ko2021} are reported in brackets. The lowest errors for models with original cutoffs are shown in bold.}
    \label{tab:single_systems}
    \vskip 0.05in
    \small
    \begin{tabular}{l rrrrrr }\toprule
        & \multicolumn{3}{c}{\textbf{Allegro}} & \multicolumn{2}{c}{\textbf{HDNN}} & \multicolumn{1}{c}{\textbf{CACE-LR}~\cite{Kim2024LearningCA} } \\
        \cmidrule(lr){2-4}\cmidrule(lr){5-6}
        & CELLI & CELLI ($J^{\mathcal R}$) & Baseline & 4G~\cite{Ko2021} & LRSR~\cite{Shaidu2024} & \\
        \midrule
        \textbf{Carbon Chains} \\ 
        \hskip 1em Energy $U$ & \textbf{0.599} & 0.609 & 0.772 &  1.194 & 1.17 & 0.73 \\
        \hskip 1em Force $F$ & \textbf{31.00} & 32.31 & 49.18 & 78.00 & 79 & 36.9 \\
        \hskip 1em Charge $Q$ & 4.003 & \textbf{3.451} & n.a. & 6.577 & 10.4 & n.a. \\
        \midrule
        \textbf{Silver Clusters} \\
        \hskip 1em Energy $U$ & 0.80 & 0.81 & 199.25 & 1.323  & 0.8 & \textbf{0.162}\\
        \hskip 1em Force $F$ & 20.33 & \textbf{20.11} & 1901.36 & 31.69 & \textbf{20} &  29.0\\
        \hskip 1em Charge $Q$& {6.360} & \textbf{1.727} & n.a. &  9.976 & 2.2 & n.a. \\
        \midrule
        \textbf{NaCl Clusters} \\
        \hskip 1em Energy $U$ &  {0.127} & \textbf{0.114} & 1.612  & 0.481 & {0.4} & 0.21\\
        \hskip 1em Force $F$ & {6.444} & \textbf{5.15} & 47.51 & 32.78 & {19} &  9.78\\
        \hskip 1em Charge $Q$ & 15.72 & \textbf{9.15} & n.a.  & 15.83 & {13.4} & n.a.\\
        \midrule
        \textbf{Gold Dimers} \\
        \hskip 1em Energy $U$ & \textbf{0.077} & \textbf{0.077} & 2.329  &  0.219 & {0.2} & (0.073) \\
        \hskip 1em Force $F$ & 12.04 & \textbf{12.01} & 123.67 &  66.00 &  {52} & (7.91)\\
        \hskip 1em Charge $Q$ & 5.510 & \textbf{4.542} & n.a. & 5.698 & {65.8} & n.a. \\
        \bottomrule
    \end{tabular}
\end{table*}

\begin{figure}
    \centering
    \includegraphics[width=\linewidth]{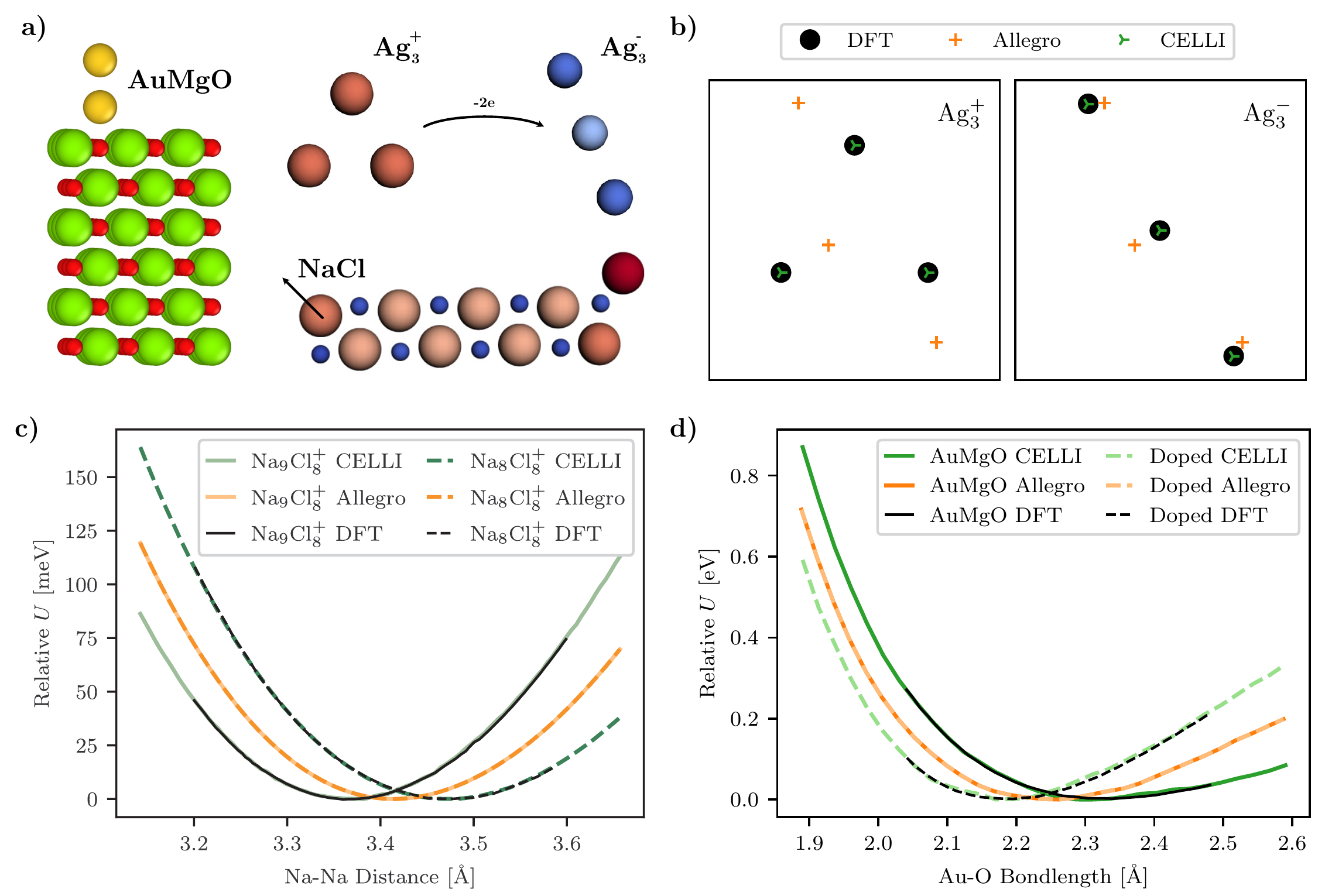}
    \caption{\textbf{Long-range and charge-dependent interactions benchmarks.} \textbf{a)} Visualization of three benchmark systems used in experiments: gold dimers on a MgO(001) surface, positively and negatively charged silver clusters, and sodium chloride clusters. In the gold dimers, colors represent atom types (Ag - yellow, Mg - green, O - red). For the other systems, colors visualize partial charges (red corresponding to positive and blue to negative charge). \textbf{(b)} Predicted minimum energy conformations for charged silver clusters. The baseline Allegro fails to distinguish between charge states, which CELLI-enhanced Allegro closely matches the DFT predicted minimum energy conformations. \textbf{(c)} Relative energies for $\mathrm{Na}_{8}\mathrm{Cl}_{8}^+$ and $\mathrm{Na}_{9}\mathrm{Cl}_{8}^+$ clusters as a function of the Na–Na distance along a predefined path (indicated by arrow). CELLI-enhanced Allegro closely reproduces DFT energy profiles and correctly identifies distinct minima for the two charge states, unlike baseline Allegro. \textbf{(d)} Predicted bond energies for a gold dimer on an MgO(001) surface in the upright (non-wetting) geometry, with and without Al doping. CELLI-enhanced Allegro matches DFT results for both cases, while baseline Allegro fails to differentiate between doped and undoped substrates.}
    \label{fig:bench_systems}
\end{figure}

In almost all cases, we observed significant improvements over the models presented by \citeauthor{Ko2021, Shaidu2024, Kim2024LearningCA} (Table \ref{tab:single_systems}).
This improvement might arise due to the use of equivariant GNNs instead of Behler-Parrinello Neural Networks utilized in 4GHDNN-based models and a learned embedding of the charge environment (For a comparison between local and environment charge embedding, see Supplementary Table~3). Moreover, in most cases --- except for the AuMgO system, where CACE-LR employed an increased cutoff --- we achieved substantial accuracy gains not only over the baseline Allegro model, where our RMSE is in some case several orders of magnitude better, but also over the non-4GHDNN-based CACE-LR model, which captures long-range effects solely through local feature augmentation. These results suggest that the improved performance stems from a novel integration of environment-dependent charges via the Qeq mechanism into the equivariant GNN framework, enabling accurate modeling of long-range interactions. Notably, the largest error reductions were achieved with CELLI when using environment-dependent hardnesses, suggesting that this variant should be preferred in future applications. However, in some cases, the improved charge predictions from the environment-dependent hardness version did not lead to substantial changes in force or energy accuracy, indicating that prioritizing charge accuracy alone may not always result in a better overall model.

In addition, we conducted experiments on these systems to verify whether the model could accurately represent long-range effects and the resulting changes in the PES.
Allegro consistently fails these tests, showing significantly higher errors and producing unphysical results (see Figure \ref{fig:bench_systems}). In contrast, CELLI yields predictions that closely match DFT calculations and align with theoretical expectations. The results confirm CELLI’s ability to model can effectively capture long-range charge transfer and electrostatics (carbon chains, NaCl clusters), handle differences between charged states (silver clusters), and accurately model energies and forces in charge-sensitive environments (gold dimers on MgO(001) surfaces). These findings underscore CELLI's strength in representing critical phenomena in diverse systems.

\paragraph{Long-range interactions for message-passing models}

To demonstrate that CELLI is applicable beyond strictly local architectures, we integrate it into the message-passing network MACE~\cite{NEURIPS2022MACE}.
Using the scalar node features $h_i^{(l)}$ and bessel radial basis edge embeddings $e_{\text{rbf},ij}$ (see Methods, Section~\ref{subsec:gnns}), we construct edge features $x^{(0)}_{ij} = (p_\text{env}(r_{ij})h_i  || e_{\text{rbf},ij})$, where $p_\text{env}$ is the envelope function of the model and $||$ is a concatenation.
The output $x^{(1)}_{ij}$ from CELLI is aggregated to a weighted residual update of the scalar node features $h_i^{(l+1)} = h_i^{(l)} + \varepsilon\sum_{j \in \mathcal N(i)} x^{(1)}_{ij}$, where $\varepsilon$ is a learnable weight for message aggregation.

We asses the effect of CELLI using the benchmarks from the previous section, excluding the silver clusters due to already beeing fully contained in the receptive field of the strictly local Allegro model.
Additionally, we compare our results to SpookyNet ~\cite{Unke2021}.

CELLI significantly enhances the performance of the baseline MACE model, in fact, in some cases it achieves errors almost ten times lower. Both SpookyNet and CELLI(6) achieve particularly low errors compared to models with two message-passing steps, likely due to their deeper architectures with six message-passing layers, which allow them to capture complex interactions even in smaller systems. However, this depth may limit their applicability to larger systems, as using many message-passing layers can become computationally impractical \cite{Musaelian2023}. 

Interestingly, while CELLI tends to produce significantly larger errors in partial charge predictions compared to SpookyNet, these discrepancies do not consistently correlate with errors in energy or force predictions. This may reflect differences in how each model utilizes charge information and is consistent with previous findings that highlight limitations of charge partitioning schemes \cite{Shaidu2024}. Overall, this comparison underscores CELLI’s effectiveness in modeling systems with non-local interactions, even within message-passing neural networks.

\begin{table*}[t]    \centering    \vskip -0.15in
    \scriptsize
    \caption{Root mean square errors (RMSE) in units of meV/atom, meV/\AA, and me, for CELLI applied to the message-passing model MACE vs. baseline MACE and SpookyNet~\cite{Unke2021}.
    The numbers of message-passing steps are given in brackets next to the model variant. Errors for models with a larger cutoff than in the original reference~\cite{Ko2021} are reported in brackets. The lowest errors for models with original cutoffs are shown in bold.
    }
    \label{tab:single_systems_mace}
    \vskip 0.05in
    \small
    \begin{tabular}{l rrrrr c}\toprule
        &  \multicolumn{3}{c}{\textbf{MACE}}  & \multicolumn{1}{c}{\textbf{SpookyNet}} \cite{Unke2021} \\
        \cmidrule(lr){2-4} \cmidrule(lr){5-5} \cmidrule(lr){6-6}
        & CELLI (6) & CELLI (2) & Baseline (2) &  (6) \\
        \midrule
        \textbf{Carbon Chains} \\ 
        \hskip 1em Energy $U$ & \textbf{0.128} & 0.398 & 0.335 & (0.364)  \\
        \hskip 1em Force $F$ & \textbf{4.36} & 21.45 & 17.68 & (5.802)  \\
        \hskip 1em Charge $Q$ & \textbf{1.273}  & 3.458 & n.a. & (0.117)  \\

        \midrule
        \textbf{NaCl Clusters} \\
        \hskip 1em Energy $U$ & 0.104 & \textbf{0.097} & 1.557 & 0.135  \\
        \hskip 1em Force $F$ & 3.92 & 3.54 & 39.96 & \textbf{1.052} \\
       \hskip 1em Charge $Q$ & 13.91 & 15.52 & n.a. & \textbf{0.111}  \\
        \midrule
        \textbf{Gold Dimers} \\
        \hskip 1em Energy $U$ & \textbf{0.065} & 0.069 & 2.13 & (0.107) \\
        \hskip 1em Force $F$ & \textbf{5.94} &  7.95 & 56.52 & (5.337)  \\
        \hskip 1em Charge $Q$ & \textbf{2.322} &  5.171 & n.a. & (1.013)  \\
         \bottomrule
     \end{tabular}
 \end{table*}

\paragraph{Generalization to chemically diverse systems}

The previous benchmarks consist of relatively small and simple systems.
Therefore, they cannot demonstrate the generalizability across a wide chemical space and the advantageous scalability of our scheme with the size of the system.
To this end, we included the OE62 dataset~\cite{OE62}, which allows us to evaluate the performance of our model on larger and more diverse systems of varying sizes, providing a complementary assessment to the simpler benchmarks focused on single structures with different charge states. In addition, we evaluate the scalability of CELLI by measuring forward-pass times across systems with varying atom counts. We compare three versions of CELLI (two using Allegro and one using MACE) against four Allegro baselines, which differ in model size and the inclusion of an additional tensor product layer. We also include various MPNN architectures, including models that incorporate other long-range correction schemes: Ewald and Neural P$^3$M ~\cite{kosmala_ewaldbased_2023, wang2024neuralp3mlongrangeinteracti} as well as baseline DimeNet++ with hyperparameters as in \citeauthor{kosmala_ewaldbased_2023} (Detailed Results in Supplementary~Table~1).

\begin{table*}[t]
    \centering
    \caption{Summary of accuracy for all trained models (cutoff 0.6 nm) on the OE62 dataset compared to other long-range modeling approaches. The \textit{Small} (S) versions of CELLI and Allegro were used to compute the benchmarks and use fewer irreps, a lower rotational order for the spherical harmonics, and a smaller hidden size for the charge-embedding networks than the \textit{Large} (L) version. The versions \textit{S+} and \textit{L+} of Allegro contain one additional Interaction Layer compared to the CELLI version.
    The number of message-passing steps of each model, if applicable, is reported in brackets behind the model name. Results for Ewald and Neural P$^3$M on DimeNet++ and PaiNN were taken directly from the references~\cite{kosmala_ewaldbased_2023, wang2024neuralp3mlongrangeinteracti}. The lowest errors are reported in bold. (*) Reported by \citeauthor{kosmala_ewaldbased_2023}.}
    \label{tab:OE62_MP}
    \vskip 0.15in
    \small
    \begin{tabular}{lccc}\toprule
    \textbf{Model} & $U$ MAE [meV] & $U$ RMSE [meV] & \# Mio. Params. \\\midrule
    \multicolumn{4}{l}{\textbf{Allegro} (--)} \\
    \hspace{1em}Baseline S           & 63.4  & 123.7 & 0.17 \\
    \hspace{1em}Baseline S+          & 60.0  & 114.5 & 0.20 \\
    \hspace{1em}Baseline L           & 61.1  & 120.9 & 0.19 \\
    \hspace{1em}Baseline L+          & 61.8  & 116.6 & 0.22 \\
    \hspace{1em}CELLI S     & 55.3  & 116.7 & 0.21 \\
    \hspace{1em}CELLI L     & 55.1  & 114.3 & 0.29 \\
    \midrule
    \multicolumn{4}{l}{\textbf{MACE} (2)}\\
    \hspace{1em}Baseline        & 48.1  & 90.1  & 2.37 \\
    \hspace{1em}CELLI       & 48.0  & \textbf{88.3} & 2.52 \\
    \midrule
    \multicolumn{4}{l}{\textbf{DimeNet++} (3)} \\
    \hspace{1em}Baseline        & 42.1 (53.8\textsuperscript{*}) & 108.4 & 2.78 \\
    \hspace{1em}Ewald \cite{kosmala_ewaldbased_2023}      & 48.1  & --    & 4.8 \\
    \hspace{1em}Neural P$^3$M \cite{wang2024neuralp3mlongrangeinteracti} & \textbf{41.5}  & --    & -- \\
    
    \midrule
    \multicolumn{4}{l}{\textbf{PaiNN} (4)} \\
    \hspace{1em}Ewald  \cite{kosmala_ewaldbased_2023} & 59.7  & --    & 15.7 \\
    \hspace{1em}Neural P$^3$M \cite{wang2024neuralp3mlongrangeinteracti} & 52.9  & --    & -- \\
    \bottomrule
    \end{tabular}
\end{table*}

\begin{figure}[t]
    \centering
    \includegraphics[width=\linewidth]{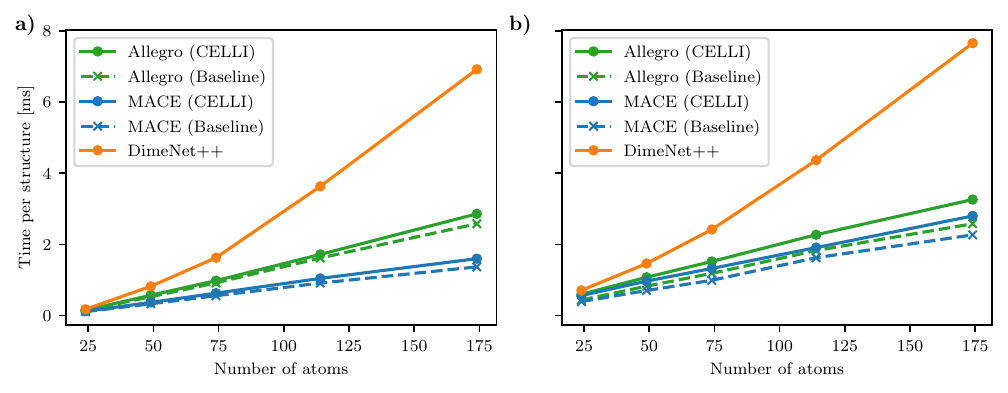}
    \caption{\textbf{Computational cost of CELLI.} Average forward pass runtime per structure for the smallest Allegro and MACE models in the baseline (B) and CELLI (C) extended variants, and Dimnet++ for different systems with different numbers of atoms. To obtain a more reliable average, several structure sizes were binned together. \textbf{a)} Shows runtime per structure for a single sample, \textbf{b)} shows the optimal runtime per structure for batch sizes [1, 10, 25, 50, 100]. Allegro and MACE exhibit significantly better performance for larger structures with a marginal impact of CELLI.}
    \label{fig:OE62}
\end{figure}

Our results show that CELLI outperformed the baseline Allegro models (Table \ref{tab:OE62_MP}) with only a marginal increase in computational cost (Figure~\ref{fig:OE62}). Notably, even the small CELLI variant outperformed the largest Allegro baseline model, demonstrating that applying the Qeq scheme is more effective than merely increasing model size, as certain effects cannot be captured without an appropriate long-range correction method. In fact introduction of CELLI to strictly local Allegro model makes its performance comparable to MPNN architectures with state-of-the-art long-range correction schemes. Moreover, CELLI combined with Allegro not only improves upon baseline models but also achieves results comparable with state-of-the-art Dimnet++ with Neural P$^3$M and significantly outperforms PaiNN models with Ewald and Neural P$^3$M corrections, which have substantially more parameters and message-passing steps.
Reducing the number of parameters and memory requirements of the models helps avoid memory-related issues \cite{fuchs2025chemtraindeployparallelscalableframework} in large-scale simulations. 
Moreover, CELLI's compatibility with a strictly local baseline model could increase its potential to scale efficiently across multiple GPUs~\cite{fuchs2025chemtraindeployparallelscalableframework}.
In the case of MACE, the improvement over the baseline is noticeably smaller, which is likely due to good performance of baseline and presence of message passing. While CELLI exhibits a slightly higher MAE than DimeNet++, it achieves a lower RMSE and substantially improved computational efficiency, reducing runtime by approximately a factor of two (Figure \ref{fig:OE62}). The discrepancy between the RMSE and MAE results could be due to more outliers in DimeNet++.  
Additionally, the MACE model has a higher potential to achieve high scalability on multi-GPU simulations than DimeNet++ and PaiNN, due to fewer message passing steps.
Thus, CELLI offers significant improvements for highly local baseline models with only a marginal increase in computational cost, which is mainly determined by the underlying architecture. Therefore, CELLI promises efficient and accurate MD simulations of large, complex structures.

It is worth noting that in both the carbon chain benchmark and the OE62 dataset, CELLI combined with Allegro performs worse than the baseline MACE model without long-range corrections. These datasets feature small organic molecules where electrostatic interactions and long range effects are not as pronounced as in other benchmark cases. Additionally, the OE62 dataset consists of ground-state geometries, further reducing the relevance of dynamic charge redistribution. Since these systems are largely dominated by local interactions, introducing message passing steps can improve the model more than CELLI, particularly when electrostatics play a limited role. Nevertheless, this case shows that CELLI can generalize to diverse chemical spaces and allows for a comparison to different models and long-range correction schemes.

\paragraph{Verifying simulation stability}

Performing MD simulations requires models to be stable for many timesteps.
To validate the robustness of CELLI, we perform a series of MD simulations at ambient conditions (see Method section).
Therefore, we train a baseline and a CELLI-enhanced Allegro on the SPICE dataset.
We selected this dataset because it provides forces for non-equilibrium low and high-energy structures.
Therefore, the dataset promotes model stability by providing much information about conformations encountered in MD simulations, compared to the OE62 dataset, which contains only minimum energy structures.

Replacing one interaction layer by CELLI reduced the energy and force mean absolute errors from $15.5$ meV/atom to $9.4$ meV/atom and from $81.4$ meV/\AA\ to $72.5$ meV/\AA.
In the MD simulations, none of the 16 selected structures suffered from instabilities such as broken bonds or overlapping particles for both Allegro variants.
Therefore, CELLI efficiently increases simulation accuracy at high efficiency for chemically diverse systems without introducing artifacts for samples unseen in training.

\section{Discussion}

This paper presents CELLI, a model-agnostic building block introducing the established Qeq method into highly descriptive equivariant GNN MLPs.
Using equivariant GNNs, CELLI can propose accurate parameters for chemically highly diverse environments.
Through the Qeq method, CELLI integrates information about long-range electrostatic interactions and charge transfer into effectively local MLPs.
Therefore, CELLI offers a solution to the long-standing challenge of accurately modeling long-range interactions with MLPs for chemically diverse systems and applications.

In a series of benchmark cases, we showed that strictly and effectively local MLPs struggle with modeling long-range electrostatic effects and charge-state dependence.
These models can effectively learn complex electrostatic environments through CELLI, significantly enhancing their predictive accuracy and physical validity.
Moreover, we showed that CELLI can generalize to chemically diverse datasets and large molecules, marginally increasing the computational costs of the baseline model.
Furthermore, in a series of molecular dynamics simulations, we demonstrated that CELLI provides robust predictions for samples unseen in training, which is crucial to running long and stable simulations.

Our method addresses crucial limitations of existing methods to model long-range interactions.
On the one hand, by leveraging highly expressive equivariant GNNs, CELLI does not rely on hand-crafted descriptors as used in Behler-Parrinello type Neural Networks~\cite{PhysRevB.83.153101}.
Thus, CELLI-enhanced models can be trained end-to-end, making the Qeq approach applicable for modeling large and complex chemical systems.
Moreover, end-to-end trained CELLI-enhanced models can learn representations for the charge environment, which is crucial to achieve state-of-the-art accuracies for strictly local GNNs.
On the other hand, it is also significantly more cost-effective and generally applicable than other proposed machine learning methods.
For example, CELLI does not require artificially defined periodicity and anisotropy as seen in lattice-based methods \cite{kosmala_ewaldbased_2023,wang2024neuralp3mlongrangeinteracti, frank2024euclideanfastattentionmachine}, but can be applied to systems with an arbitrary number of periodic dimensions.
Additionally, CELLI can be applied to strictly local GNNs, which are highly parallelizable across multiple GPUs~\cite{fuchs2025chemtraindeployparallelscalableframework}.
Therefore, CELLI's flexibility, in combination with high accuracy, generalizability, and efficiency, makes it ideal to run large-scale and accurate molecular dynamics simulations of complex systems under strict computational cost constraints.

In the feature, we plan to interface CELLI with the large-scale molecular dynamics simulation framework LAMMPS~\cite{thompsonLAMMPSFlexibleSimulation2022}. LAMMPS provides efficient algorithms for charge equilibration and enables running molecular dynamics simulations in parallel on multiple GPUs.
Therefore, this integration would simplify deploying CELLI to large-scale simulations.
Moreover, we plan to extend CELLI with other physics-based priors~\cite{thurlemannRegularizedPhysicsGraph2023,kabyldaMolecularSimulationsPretrained2025a,Anstine2023},
which might further reduce its costs while increasing accuracy and robustness.
Finally, we plan to assess CELLI's capabilities in predicting simulation observables, such as IR spectra in vacuum~\cite{gasteggerMachineLearningMolecular2017} and under electric fields~\cite{gasteggerMachineLearningSolvent2021}, which require accurate modeling of dynamics and electrostatics.

\section{Methods}

\subsection{Graph Neural Networks}
\label{subsec:gnns}

Molecular systems can be represented as graphs by describing atoms as nodes and defining edges between neighboring atoms within a fixed cutoff radius, allowing GNNs to learn atom-centered representations.
In the first step, GNNs embed this graph, assigning initial features $\bm h_i^{0}$ to the nodes and features $\bm x_{ij}$ to the edges from atom species $Z_i$ and atom displacements.
Subsequently, GNNs encode the graph by iteratively updating edge and node features that are finally read out to obtain node, edge, and graph property predictions.

The popular class of Message-passing neural networks (MPNNs) class, first formalized by \citeauthor{gilmerMPGNN}, encodes the graph by iteratively performing message-passing
\begin{align}
    \bm m_i^{l+1} & = \sum_{j\in \mathcal N(i)} \mathcal M^{l}(\bm h_i^l, \bm h_j^l, \bm x_{ij}),\\
    \bm h_i^{l+1} & = \mathcal U^{l}(\bm m_i^{l+1}, \bm h_i^l),
\end{align}
where $\mathcal M^l$ and $\mathcal U^l$ are learnable functions of the layer $l$.
As messages $\bm m_i^l$ contain information from all graph neighbors $j \in \mathcal N(i)$ of a particle $i$, MPNNs pass information of each atom's neighborhood along the graph.
Therefore, message-passing gradually expands the atom's receptive field and enables the capture of many-body correlations \cite{Musaelian2023, schütt2021equivariantmessagepassingprediction}.

Unfortunately, this information propagation complicates parallelized implementations of GNNs, e.g., in large-scale atomistic MD frameworks such as LAMMPS \cite{thompsonLAMMPSFlexibleSimulation2022}. 
Therefore, strictly local architectures such as Allegro \cite{Musaelian2023} have been proposed.
Conceptually, Allegro updates the directed edge features through the steps
\begin{align}
    \bm w_{ij}^{l+1} & = \sum_{k\in \mathcal N(i)} \mathcal W^l\left(\bm x_{ij}^l, \bm x_{ik}^l\right), \\
    \bm x_{ij}^{l+1} & = \mathcal U^l \left(\bm w_{ij}^{l+1}, \bm x_{ij}^{l+1}\right),
\end{align}
where $\bm w_{ij}^{l+1}$ contains information from all edges that originate from the same node.
Corresponding to the message-passing framework, the function $\mathcal W^l$ encodes information about the environment of an edge into the update function $\mathcal U^l$.
However, as two directed edges between nodes can contain different information ($\bm x^l_{ij} \neq \bm x^l_{ji}$), no information is passed along the graph. 

\subsection{Efficient Computation of Electrostatic Interactions}
\label{subsec:longrange}

Electrostatic effects are commonly approximated by coulombic interactions.
For a system of $N$ charges $Q$ with Gaussian density, located at the centers of the particles $\bm R$, the coulombic interaction potential is
\begin{align}
    U_\text{Coul}(\bm R, \bm Q) = \sum_{i}^N\sum_{j>i}^N\frac{\operatorname{erf}(\alpha_{ij} r_{ij})}{r_{ij}}Q_iQ_j + \sum_{i=1}^N\frac{2\alpha_{ii}}{\sqrt{\pi}}Q_i^2,\label{eq:coulomb}
\end{align}
where $\alpha_{ij} = \frac{1}{\sqrt{2}}(\gamma_i^2 + \gamma_j^2)^{-1/2}$ depends on the radii $\gamma_i$ of the charges separated by a distance $r_{ij} = \lVert|\bm R_i - \bm R_j\rVert$ \cite{CENT}. 
These interactions can extend over larger distances as the interaction decays approximately with the factor $1 / r$.
Moreover, the contributions from distant charges must be accurately captured without truncation or oversimplification \cite{Anstine2023}.
Therefore, coulombic interactions are more challenging to model efficiently than, e.g., short-ranged van-der-Waals interactions.

Nevertheless, classical approaches have been proposed to model long-range interactions efficiently without computing direct pairwise interactions beyond a small cutoff.
Essentially, these methods decompose the interaction potential into a rapidly decaying short-range part and a smooth but slowly decaying long-range part.
The methods then treat the short-range part directly like other short-range interactions.
However, as the long-range part still accounts for contributions from distant charges, a more efficient computation requires a different treatment.
For example, the Fast Multipole Method \cite{greengardFastAlgorithmParticle1987} hierarchically groups particles and computes distant interactions between these clusters collectively to achieve a $O(N)$ scaling with respect to the number of particles.
Especially for periodic systems, the Smooth Particle Mesh Ewald (SPME) method \cite{essmannSmoothParticleMesh1995} computes long-ranged interactions more efficiently in the reciprocal space. Similar to the short-ranged part in real space, the long-ranged part decays quickly in the reciprocal space and can be truncated without losing accuracy.
Additionally, by mapping charges to a grid leveraging B-spline interpolation for smooth gradients and employing fast Fourier transforms, it achieves a computational complexity of $O(N\log N)$. Notably, SPME is not limited to periodic systems but can be generalized to systems with partial or fully non-periodic boundary conditions, e.g., to treat isolated clusters \cite{martynaReciprocalSpaceBased1999}.

\subsection{Charge Equilibration Method (Qeq)}

\label{subsec:qeq_method}

Several approaches can compute long-ranged electrostatic interactions accurately and efficiently in many systems.
Nevertheless, these interactions must be adequately parametrized for the respective systems by assigning partial charges to the atoms.
Assigning fixed partial charges can introduce significant errors due to charge transfer induced by changes in the chemical environment \cite{CENT}.
Therefore, methods with dynamic partial charge assignment are necessary to accurately model molecular systems with significant electrostatic interactions.

To model environment-dependent partial charges, the Charge Equilibration (Qeq) method \cite{qeq} proposes to redistribute charges in the system to minimize the total energy while maintaining charge conservation $\sum_{i=1}^N Q_i = Q_\text{tot}$.
In the Qeq method, the contribution of charges to the total energy
\begin{align}
    U_\text{Qeq}(\bm R, \bm Q) = U_{\text{Coul}}(\bm R, \bm Q) + \sum_{i=1}^N\left[ \chi_i Q_i + \frac{J_{ii}}{2} Q_i^{2} \right]
\end{align}
consists of the coulombic interaction between charges $U_\text{Coul}$ given in equation~\eqref{eq:coulomb} and a second-order approximation of the charge-core interaction determined by the electronegativities $\chi_i$ and chemical hardnesses $J_i$.
Due to the form of the coulombic interaction, the charge energy is quadratic in $\bm Q$.
Consequently, the minimum of $U_\text{Qeq}$ is the solution of the linear system
\begin{align}
    \left[\left.\frac{\partial^2 U_\text{Coul}}{\partial Q_i\partial Q_j}\right|_{\bm R} + J_{ii} \right] Q_j = -\chi_i, 
\end{align}
subject to the charge conserving equality constraint $\bm 1^T\bm Q = Q_\text{tot}$.
For smaller systems, direct linear solvers can determine the optimal charges within a short runtime.
However, due to the cubic scaling with the number of particles $O(N^3)$, several other approaches have been proposed to solve the system in quadratic \cite{nakanoParallelMultilevelPreconditioned1997} or quasi-linear time \cite{Gubler2024},
leveraging efficient treatments of long-range interactions outlined in section~\ref{subsec:longrange}.

\subsection{Systems and Datasets}
\label{subsec:systems}

\paragraph{Benchmarks for long-range and electrostatic interactions}

The benchmark datasets for long-range and electrostatic interactions comprise four organic and inorganic systems with up to four different species in free and periodic boundary conditions~\cite{Ko2021}.
For each system, DFT computations of energies and forces were obtained with the PBE functional, while charges were generated with Hirshfeld population analysis. The datasets are available at \url{https://doi.org/10.24435/materialscloud:f3-yh}.

\begin{description}
    \item[Carbon Chains]~\\ The first benchmark system consists of neutral and charged carbon chains. C$_{10}$H$_2$ is a neutral linear chain of carbons terminated with hydrogen atoms, while C$_{10}$H$_3^+$ is obtained by protonating one end of the chain, leading to global charge redistribution. This system highlights how a given model accounts for long-range charge transfer caused by local perturbations.\\

    \item[Silver Clusters]~\\ The second benchmark involves triangular and linear silver trimers (Ag$_3$) with total charges of +1 and -1, respectively. These systems test the model's ability to handle differences in charge states, geometries, and identification of energetically favourable conformations.\\

    \item[Sodium Chloride Clusters]~\\ We also evaluated the sodium chloride clusters benchmark, consisting of Na$_8$Cl$_8^+$ and Na$_9$Cl$_8^+$. In these systems, moving a sodium atom along a predefined path reveals two distinct energy minima, which are sensitive to long-range electrostatics and charge redistribution and can demonstrate the model's ability to accurately predict changes in the potential energy surface.\\

    \item[Gold Dimers on MgO(001) surface]~\\ The final benchmark is a periodic system consisting of a gold dimer (Au$_2$) adsorbed on MgO(001) surfaces, both undoped and Al-doped. Two configurations were considered: “wetting,” where both Au atoms lie near Mg atoms, and “non-wetting,” where one Au atom binds to an O atom while the other remains farther away. These configurations assess the model's ability to capture adsorption energies and forces in charge-sensitive periodic settings.
\end{description}

\paragraph{OE62 dataset}

The OE62 dataset provides a diverse benchmark for the evaluation of our model, as it consists of 62,000 organic molecules extracted from the Cambridge Structural Database, with DFT-optimized geometries at the PBE level, including van der Waals corrections \cite{OE62}. OE62 spans a broad chemical space, with up to 174 atoms and 16 elements, offering a comprehensive test case for assessing the scalability and generalizability of models on chemically complex and diverse systems.
The dataset is available at \url{https://doi.org/10.14459/2019mp1507656}.

\paragraph{SPICE dataset}

The SPICE dataset~\cite{eastman2023SPICEv2} (v2.0.1) spans a large chemical space of peptides and drug-like molecules consisting of 17 different chemical species, including low and high energy conformations and systems with non-zero net charge.
Each sample provides DFT computed energies and forces using the $\omega\mathrm{B97M-D3(BJ)}$ functional with dispersion correction and the def2-TZVPPD basis set as well as MBIS charges.
For this paper, we selected the \emph{Amino Acid Ligand, PubChem Sets, DES370K, DES Monomers, and Dipeptides} subsets.
The full dataset is available at \url{https://doi.org/10.5281/zenodo.10975225}.

\subsection{Model optimization}

We performed all experiments in the deep-learning framework JAX using chemtrain \cite{fuchs2024chemtrainlearningdeeppotential} to train the models.
Therefore, we adapted JAX-MD \cite{schoenholzJAXFrameworkDifferentiable2021}, and JAX compatible implementations of Allegro \cite{marioAllegrojax} and DimeNet++ \cite{ thalerLearningNeuralNetwork2021, gasteigerFastUncertaintyAwareDirectional2022}, and MACE~\cite{NEURIPS2022MACE,marioMACEjax}.

\paragraph{Preparation of energies}

The reference energies in the datasets contain large negative shifts.
Therefore, we shift the reference energies $U$ by species-dependent constant shifts $U_s$ to obtain the target energies
\begin{align}
    \hat{U}_i = U_i - \sum_{s=1}^SU_{s}N_{s,i}
\end{align}
where $N_{s,i}$ counts the occurrences of species $s$ in sample $i$.
We determined the shifts $U_s$ through a ridge-regression fit to the dataset.

\paragraph{Training}

We train the models via the Force Matching method \cite{ercolessiInteratomicPotentialsFirstPrinciples1994, fuchs2024chemtrainlearningdeeppotential}.
Therefore, we optimize the parameters $\theta$ to minimize the loss function
\begin{align}
        \mathcal L(\theta) = \frac{1}{D}\sum_{i=1}^D\biggl[&\gamma_U\lVert U_\theta(\bm R_i) - \hat U_i\rVert^2
        +\frac{\gamma_F}{3N_i}\lVert \bm F_\theta(\bm R_i) - \hat{\bm F}\rVert^2
        +\frac{\gamma_Q}{N_i}\lVert \bm Q_\theta(\bm R_i) - \hat{\bm Q}\rVert^2\biggr]\label{eq:loss}
\end{align}
between the reference values $\hat U, \hat{\bm F}, \hat{\bm Q}$ and the model predictions $U, \bm F, \bm Q$ for $D$ samples $\bm R$ of the training dataset via stochastic optimization using the ADAM optimizer \cite{kingmaAdamMethodStochastic2017} and a polynomial step-size schedule with weight decay.
The parameters $\gamma_U, \gamma_F, \gamma_Q$ balance the contributions of the targets to the loss and are set problem-specific.
We monitor the convergence by empirically estimating the loss on a disjoint validation split and select the parametrization $\theta$ that yielded the lowest error on the validation split.

\paragraph{Hyperparameters}

Model cutoffs were chosen similar to \citeauthor{Ko2021} for the four benchmark systems and to \citeauthor{kosmala_ewaldbased_2023} for the OE62 dataset (Supplementary~Table~2). In the four benchmark systems and for the SPICE dataset, CELLI is replaced by an additional interaction layer to obtain the baseline Allegro model. For the OE62 dataset, CELLI is excluded without replacement (S, L) or replaced by an additional Interaction Layer (S+, L+) to obtain the Allegro baseline variants. For the MACE model, CELLI is always excluded without replacement. DimeNet++ hyperparameters are similar to~\citeauthor{kosmala_ewaldbased_2023}, except for the loss function, which is chosen to comply with Equation~\ref{eq:loss}.

\subsection{Benchmarks on OE62 and SPICE datasets}

\paragraph{Timing OE62 Forward Passes}

We evaluate the forward pass run times for all models on a single \textit{NVIDIA~A100}.
For evaluating the computational performance, we partition the training split at $[25, 50, 75, 115, 174]$ atoms per molecule and choose the 5000 largest structures.
For each subset, we choose the maximum number of edges and triplets to account for the maximum required by any sample in the subset.
We then time the forward pass for batch sizes of $[1, 10, 25, 50, 100]$ for the ahead-of-time compiled model.

\paragraph{Simulating SPICE systems}

We run MD simulations for 16 different systems drawn equally from the four different subsets \textit{PubChem Sets}, \textit{Amino Acid Ligands}, \textit{Dipeptides}, and \textit{DES370K}.
Starting from a randomly selected conformation from the testing set, we run simulations for $1$ ns with a step size of $0.5$ fs at $300$ K using a stochastic thermostat~\cite{gogaEfficientAlgorithmsLangevin2012}  with a friction coefficient of $100\ \text{ps}^{-1}$.
Therefore, we perform $2$ million update steps per model.

\subsection*{Data Availability}

The datasets used in this study are publicly available to download (see Methods).

\subsection*{Code Availability}
The software \texttt{chemtrain} used to train the models and perform MD simulations is publicly available at \url{https://github.com/tummfm/chemtrain}.
Adapted models, training, and evaluation scripts are not publicly available but may be made available to qualified researchers on reasonable request from the corresponding author.

\subsection*{Author Contribution Statement}

P.F. and J.Z. conceptualized the study. P.F. developed the model and implemented the software. P.F. and M.S. performed the experiments and wrote the manuscript. M.S. analyzed the results. J.Z. supervised the project, reviewed the manuscript, provided resources, and acquired funding.

\subsection*{Acknowledgments}
Funded by the European Union. Views and opinions expressed are however
those of the author(s) only and do not necessarily reflect those of the European Union or the
European Research Council Executive Agency. Neither the European Union nor the granting
authority can be held responsible for them.
This work was funded by the ERC (StG SupraModel) - 101077842 and the Deutsche Forschungsgemeinschaft (DFG, German Research Foundation) - 534045056 and 561190767.

\clearpage

\begin{appendices}

\end{appendices}

\bibliography{sn-bibliography}

\end{document}